# TSViT: A Time Series Vision Transformer for Fault Diagnosis

Shouhua Zhang, Jiehan Zhou, Xue Ma, Susanna Pirttikangas, Chunsheng Yang

*Abstract*—Traditional fault diagnosis methods using Convolutional Neural Networks (CNNs) often struggle with capturing the temporal dynamics of vibration signals. To overcome this, the application of Transformer-based Vision Transformer (ViT) methods to fault diagnosis is gaining attraction. Nonetheless, these methods typically require extensive preprocessing, which increases computational complexity, potentially reducing the efficiency of the diagnosis process. Addressing this gap, this paper presents the Time Series Vision Transformer (TSViT), tailored for effective fault diagnosis. TSViT incorporates a convolutional layer to extract local features from vibration signals, alongside a transformer encoder to discern long-term temporal patterns. A thorough experimental comparison on three diverse datasets demonstrates TSViT's effectiveness and adaptability. Moreover, the paper delves into the influence of hyperparameter tuning on the model's performance, computational demand, and parameter count. Remarkably, TSViT achieves an unprecedented 100% average accuracy on two test sets and 99.99% on another, showcasing its exceptional diagnostic capabilities.

## 1 Introduction

Rotating machinery is an important component for modern industrial equipment. With the increasing enhancement of product complexity and the integration of functional modules, it poses greater challenges to the security, stability, and overall equipment's robustness [1]. Consequently, the development of effective fault diagnosis methods for rotating machinery becomes imperative.

As Industrial Internet of Things (IIoT) and big data analytics advance, research on fault diagnosis based on deep learning has emerged [2]. Deep learning models exhibit robust learning capabilities, allowing for the automatic extraction of fault features from data without manual intervention. This significantly reduces the reliance on expert experience and domain knowledge.

Many deep learning models have been successfully proposed for fault diagnosis in recent years such as Convolutional Neural Networks (CNNs) [3][4], Recurrent Neural Networks (RNNs) [5][6], Deep AutoEncoders (DAEs) [7] [8], and Deep Belief Networks (DBNs) [9][10]. However, the sequential nature of RNN models poses challenges in terms of training parallelization, gradient explosion or vanishing, and is susceptible to significant long-term memory loss. Convolutional filters in CNNs are constrained to process a small local region and can not capture global information adequately [11]. Unsupervised DAEs and DBNs exhibit unsatisfactory performance in the case of large-scale and complex datasets due to their structural characteristics [12].

Transformer has achieved significant improvement in Natural Language Processing (NLP) since it was first proposed based on the self-attention mechanism [13]. However, its ability to capture local features is insufficient. Inspired by transformer models in NLP, researchers recently started to apply transformers to Computer Vision (CV) tasks, known as vision transformers, achieving notable results [14]. A vision transformer divides an input image into a sequence of patches. Each patch is then converted into a vector and reduced to a smaller dimension. These vector embeddings are subsequently processed by a transformer. Up until now, there has been limited research on utilizing vision transformers for fault diagnosis.

Motivated by these observations, we developed a new fault diagnosis method based on vision transformer for rotating machinery, namely TSViT. On one hand, it incorporates a convolutional layer to divide vibration signals into patches and capture local features. On the other hand, it utilizes a transformer encoder to learn long-term temporal information. This enables comprehensive spatiotemporal feature extraction. The main contributions of this paper are as follows.

1) Proposed TSViT, a Time Series Vision Transformer model for fault diagnosis with the capability to directly process raw time series signals.

2) Developed a time series patch embedding method to enable TSViT to accept time domain signals in either one-

*Corresponding author: Jiehan Zhou.*

S. Zhang is with ITEE, University of Oulu, Oulu, Finland (e-mail: shouhua.zhang@oulu.fi; susanna.pirttikangas@oulu.fi).

J. Zhou is with Shandong University of Science and Technology, Qingdao, China and Docent at University of Oulu, Oulu, Finland (e-mail: jiehan.zhou@ieee.org).

X. Ma is with Department of Automation, Tsinghua University, Beijing, China (e-mail: xuema1992@163.com).

S. Pirttikangas is with ITEE, University of Oulu, Oulu, Finland (e-mail: susanna.pirttikangas@oulu.fi).

C. Yang is with Institute of Artificial Intelligence, Guanzhou University, Guangzhou, China (e-mail: yang400@hotmail.com).

dimensional or multi-dimensional formats as its input instead of image data.

3) Designed the experiments with three distinct datasets. The results demonstrate that TSViT can achieve highly accurate fault diagnosis without using any preprocessing techniques.

The remainder of this paper is organized as follows. Section 2 reviews the works related to vision transformer for fault diagnosis. Section 3 presents the framework of TSViT. Section 4 presents the experiments and the datasets. Section 5 validates the effectiveness and the superiority of TSViT over other methods with a comparative analysis of its hyperparameters' impact on model performance, computational complexity, and overall parameter quantity. Section 6 draws the conclusion and outlines future work.

## 2 RELATED WORK

Parmar and Vaswani et al. [15] from Google Brain first applied the transformer to CV in 2018 and proposed the Image Transformer model. Carion and Massa et al. [16] from Facebook AI proposed a method DERT for object detection that uses a transformer architecture to directly predict a set of bounding boxes for each object in an image. Dosovitskiy and Beyer et al. [17] from Google Brain proposed ViT model that applies the transformer architecture to sequences of image patches. The success of Image Transformer, DERT, ViT have significantly propelled the rapid development of vision transformers and swept the entire CV field. In the latest survey on vision transformer, Han and Wang et al. [18] acknowledged that nowadays transformer is a potential alternative to CNN. The successful application of transformers implies a trend where the transformer architecture is becoming a unified framework for developing models in both CV and NLP. The adoption of transformer facilitates the seamless integration of vision and language modeling. This emerging trend eases the joint modeling for vision and language processing, fostering a shared learning experience that accelerates advancement in both domains.

Few studies were reported in applying vision transformers to fault diagnosis recently. Weng et al. [19] proposed a one-dimensional vision transformer with multi-scale convolution fusion (MCF-1DViT), which combines CNN and vision transformer to diagnose rolling bearing faults based on vibration signals. They designed a multi-scale convolution fusion layer to capture fault features in multiple time scales from vibration signals before applying the transformer. Tang et al. [20] proposed an integrated vision transformer model based on wavelet transform for bearing fault diagnosis. It utilizes the discrete wavelet transform and continuous wavelet transform to generate time–frequency diagrams with vibration signals before applying vision transformer. He et al. [21] converted time series signals into time-frequency diagrams with Short-Time Fourier Transform (STFT) and combined the Siamese network and vision transformer to propose the Siamese vision Transformer (SviT). SviT aims to extract feature vectors in a high-level space for bearing fault diagnosis. Zim et al. [22] used STFT to convert vibration signals to two-dimensional time-frequency diagrams and fed these images into a vision transformer for bearing fault diagnosis. Their resulted performance shows that vision transformer proves to be highly promising for fault diagnosis. Nevertheless, these approaches demand extensive preprocessing of vibration signals prior to employing the transformer, diminishing the computational efficiency of data processing.

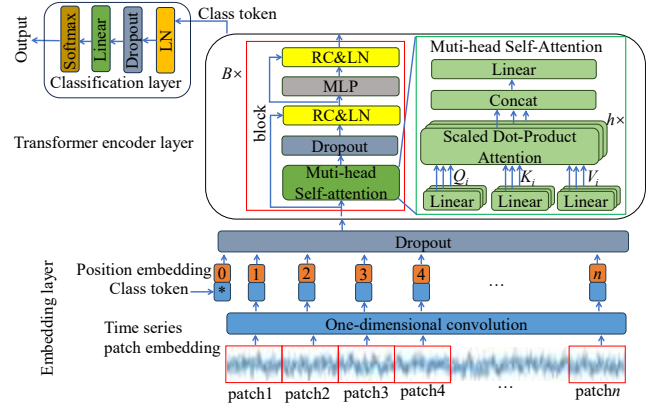

Fig.1. The framework of TSViT

## 3 TIME SERIES VISION TRANSFORMER

Figure 1 presents the proposed TSViT model. It employs a convolutional layer to extract local features and employs the transformer encoder to fully learn the long-term temporal features of input signals. It addresses the issue of a restricted receptive field in convolutional kernels, enabling to simultaneously capture long-term dependencies in signals.

### 3.1 Embedding layer

The embedding layer consists of time series patch embedding, class token, and position embedding.

#### 3.1.1 Time series patch embedding

Currently, employing vibration signals remains a prominent approach for diagnosing faults in rotating machinery [23]. Raw vibration signals typically exist in the form of one-dimensional or three-dimensional time series data. In contrast, vision transformers usually process 3-channel RGB images. The utilization of time series patch embedding empowers TSViT model to directly handle time series data in both one-dimensional and multi-dimensional formats, capturing local features effectively. The basic principle of time series patch embedding is explained as follows.

A time series vibration signal can be denoted as $X \in R^{L \times C}$, where $L$ is the length of the vibration signal, and $C$ is the dimension of the vibration signal. First, we split the input time series data into fixed-sized patches, similar to image segmentation in ViT [17]. Each patch can be denoted as $p \in R^{L_p \times C}$, where $L_p$ is the length of each patch. We can get $L = L_p \times n$, where $n$ is the number of patches. Afterward, convolution is applied to each patch, as shown in Eq. (1). The size of the convolution kernel is the same as that of $p$, and the stride is $L_p$.

$$e_p^k = \sum_{i=0}^{L_p-1} \sum_{j=0}^{C-1} p(i,j) \times weight^k(i,j) + b^k,$$
$$1 \leq k \leq m \quad (1)$$

where $weight^k$ indicates the weight of the $k$-th convolution kernel and $b^k$ indicates the bias of the $k$-th convolution kernel. $m$ is the number of convolution kernels and the dimension of the time series patch embedding. $e_p$ represents the embedding result of patch $p$, as depicted in Eq. (2). The result of patch embedding on $X$ is as shown in Eq. (3). Here, the utilized convolution is one-dimensional, implying that the convolution operates in a single direction. It's important to note that one-dimensional convolution refers to the dimensionality of the convolution's direction, not the dimensionality of the input data.

$$e_p = [e_{p,1}, e_{p,2}, \cdots, e_{p,m}] \quad (2)$$
$$PE(X) = \{e_1, e_2, \ldots, e_n\} \quad (3)$$

### 3.1.2 Class token

The class token is a special marker employed here to classify the entire input sequence. TSViT model introduces a learnable class token embedding, which is added to the front of the sequence of embedded patches, similar to the approach used in BERT model [24]. The final hidden state corresponding to this class token is used as the representation for classification. The class token is denoted as $e_0 \in R^{1 \times m}$ and the sequence of embedded patches including the class token is as shown in Eq. (4). $e_0$ will connect to all the embedded patches after the multi-head self-attention mechanism and fuse the features from all the embedded patches. Therefore, $e_0$ can be used as the feature map.

$$PEC(X) = \{e_0, PE(X)\}$$
$$= \{e_0, e_1, e_2, \ldots, e_n\} \quad (4)$$

### 3.1.3 Position embedding

As the multi-head self-attention does not account for position information in an input sequence, impacting the capture of relationships between data, TSViT incorporates learnable position embeddings. These position embeddings, represented as learnable position vectors, are added to patch embeddings to preserve position information. The position embedding on the sequence of embedded patches with the class token can be denoted by Eq. (5), where $v_i \in R^{1 \times m}$ ($i = 0,1, \ldots, n$).

$$Eembedding(X) = PEC(X) + \{v_0, v_1, v_2, \ldots, v_n\}$$
$$= \{e_0, e_1, e_2, \ldots, e_n\} + \{v_0, v_1, v_2, \ldots, v_n\} \quad (5)$$

There are two formats for position embedding: one-dimension and two-dimension. According to the experiments conducted on ViT [17], it was indicated that incorporating a two-dimensional position embedding does not result in significant performance enhancements. In this case, TSViT employs the standard learnable one-dimensional position embedding. Consequently, it employs a dropout for enhancing its generalization capability.

### 3.2 Transformer encoder layer

The transformer encoder layer consists of sequential stacking of basic blocks, each sharing the same structure. Each block consists of a Multi-head Self-Attention (MSA) layer, a MultiLayer Perception (MLP) layer, a Residual Connection (RC) layer, and a Layer Normalization (LN) layer. The output of each block serves as the direct input for the subsequent block.

### 3.2.1 Muti-head self-attention

The multi-head self-attention mechanism contributes to establishing long-term dependencies between input sequences by concatenating and fusing the outputs of multiple independent single-head self-attentions using learnable parameters. A single self-attention is a mechanism used in deep learning, which allows a model to weigh different parts of the input sequence when making predictions. The multi-head self-attention enables the model to simultaneously focus on different aspects of the input, thereby enhancing its learning expressiveness and generalization ability.

The self-attention defines three learnable weight matrices: $W^Q \in R^{m \times d_k}$, $W^K \in R^{m \times d_k}$, and $W^V \in R^{m \times d_v}$. Let $Y \in R^{n \times m}$ represent a sequence containing $n$ embedded patches, where $m$ is the dimensionality of each embedded patch. $Y$ multiplies $W^Q$, $W^K$, and $W^V$, respectively to obtain the query matrix $Q$, key matrix $K$, and value matrix $V$, as shown in Eq. (6).

$$Q = YW^Q, K = YW^K, V = YW^V \quad (6)$$

The output of the self-attention is the weighted sum of $V$, and the corresponding weight matrix can be calculated in various ways. Among them, the scaled dot-product attention is simple and easy to parallelize and does not introduce additional parameters into the model. It has been widely used and its specific calculation is shown in Eq. (7).

$$Attention(Q, K, V) = softmax\left(\frac{QK^T}{\sqrt{d_k}}\right)V \quad (7)$$

where $QK^T$ is attention score, $d_k$ is the dimensionality of $Q$ and $K$, and $\sqrt{d_k}$ is a scaling factor. The scaling factor can avoid gradient instability when $d_k$ is large.

The single-head attention mechanism is limited by the feature space, and its modeling ability is difficult to satisfy the various complex relationships that may exist between data. Hence, the multi-headed attention is needed. There are multiple independent self-attention heads in the muti-head self-attention mechanism. Each head has its own learnable weight matrix $W_i^Q \in R^{m \times d_{k,i}}$, $W_i^K \in R^{m \times d_{k,i}}$, $W_i^V \in R^{m \times d_{v,i}}$. The calculation process is shown in Eq. (8)-(10).

$$Q_i = YW_i^Q, K_i = YW_i^K, V_i = YW_i^V, i = (1,2, \ldots, h) \quad (8)$$
$$Head_i = Attention(Q_i, K_i, V_i), i = (1,2, \ldots, h) \quad (9)$$
$$MultiHead(Q, K, V) = Concat(Head_1, Head_2, \ldots, Head_h)W^O \quad (10)$$

where $W^O \in R^{hd_{v,i} \times m}$, $h$ is the number of Head, $d_{v,i}$ is the dimensionality of $V_i$.

$Q_i, K_i, V_i$ can be regarded as the split of $Q, K, V$ in single-head self-attention under different feature subspaces. The multi-head self-attention mechanism extracts the correlation between features from multiple angles and merges the information extracted by each self-attention head to obtain richer and more comprehensive feature information.

A dropout, RC and LN are added after MSA as shown in Fig. 1. The output after these operations in $l$th block can be described as Eq. (11).
$$z_l^{MLR} = LN(dropout(MSA(z_{l-1})) + z_{l-1}), l = 1,2,\ldots,B \quad (11)$$
where $B$ is the number of the basic blocks in the transformer encoder and $z_{l-1}$ is the output of the $(l-1)$th block.

### 3.2.2 Multilayer perceptron

MLP consists of two linear transformation layers, a nonlinear activation function between them, and two dropout layers. This structure is a classic method for feature extraction. The output of MLP in $l$th block can be described as Eq. (12).
$$z_l^{MLP} = dropout(dropout(\sigma(z_l^{MLR}W_1^l + b_1^l))W_2^l + b_2^l) \quad (12)$$
where $W_1^l \in R^{m \times d_{MLP}}$, $b_1^l \in R^{d_{MLP}}$, $W_2^l \in R^{d_{MLP} \times m}$, $b_2^l \in R^m$, $\sigma$ is the activation function. $d_{MLP}$ is the embedding dimensionality of the nonlinear transformation in MLP. Gaussian Error Linear Unit (GELU) is used as the activation function in MLP. A RC and LN are added after MLP as shown in Fig. 1. The output of $l$th block can be described as Eq. (13).
$$z_l = LN(z_l^{MLP} + z_l^{MLR}) \quad (13)$$

### 3.3 Classification layer

TSViT introduces the classification layer to transform the feature map extracted by the transformer encoder into one-hot encoding for pattern recognition. The classification layer consists of an LN, a dropout and a linear transformation as shown in Fig. 1. The calculation process is shown in Eq. (14).

The entire output of the transformer encoder is not used as the input to the classification layer. The input is the class token of the last block in the transformer encoder and can be denoted as $z_B^0$. It is the extracted feature map from the input vibration signal as well.
$$Class(z_B^0) = softmax(Dropout(LN(z_B^0))W_{class} + b_{class}) \quad (14)$$
where $W_{class} \in R^{m \times N_c}$, $b_{class} \in R^{N_c}$, and $N_c$ is the number of categories.

The cross-entropy is the loss function in this model which is commonly used.

## 4 DATASETS

We tested TSViT with three datasets to verify its effectiveness.

1) PBR dataset. Figure 2(a) presents the experimental environment used for collecting PBR dataset. The vibration signals were collected by vibration acceleration sensors. There are three types of faults: pedestal looseness (PL), broken blade of fan (BBF), and rotor unbalance (RU). The rotational speed was set to 1500 r/m, and the sampling frequency was set to 1280 Hz. The sampling duration was set to 8 seconds each time. 10240 sensor data were collected each time. The device was sampled 100 times separately under three fault types and under the normal condition (NC). We resampled the data using a sliding window without any overlap. Non-overlapping partitions avoid test leakage and guarantee fair comparisons [25]. The width of the sliding window is 2048, resulting in a total of 2000 samples, with 500 samples for each type. 80% of samples from each type are randomly chosen for the training set, while the remaining 20% are for the test set. The details are listed in Table 1.

Table 1. PBR dataset

| Types | Marks | Label | Training | Test |
|---|---|---|---|---|
| Normal condition | NC | 0 | 400 | 100 |
| Pedestal looseness | F1 | 1 | 400 | 100 |
| Broken blade of fan | F2 | 2 | 400 | 100 |
| Rotor unbalance | F3 | 3 | 400 | 100 |

2) CWRU dataset. CWRU dataset [26] is widely used in rotating machinery fault diagnosis. In this paper, we applied it to validate the effectiveness and generalization capability of TSViT. Figure 2(b) presents the experimental bench for collecting CWRU dataset. The experimental bearing model at the drive end was 6205-2RS. Two one-way acceleration sensors were installed at the drive end and fan end to measure the vibration signals in different fault conditions. The entire rotating shaft was powered by a 2 horsepower (HP) motor, and the motor imposed varying loads on the rotating shaft. A power meter and a torque sensor were added to the rotating shaft to detect the operating status of the motor in real time.

The experiment involved nine types of bearing faults, specifically single-point faults. These faults had a depth of 0.11 inches and were implanted at the inner raceway, rolling ball, and outer raceway. The diameters of these faults were 0.007, 0.014, and 0.021 inches, respectively, and the implantation was done using electro-discharge machining. When the experimental load was 0HP, 1HP, 2HP, and 3HP, the corresponding rotation speeds were 1797 r/m, 1772 r/m, 1750 r/m, and 1730 r/m, respectively. The experiment also collected vibration signals under the normal condition (NC) with the mentioned load conditions. The sampling frequency was set to 12kHz. The vibration signal data for each state in CRWU dataset consists of hundreds of thousands sampling points. Therefore, it cannot be directly used for training and testing the model. We resampled it using a sliding window with a width of 2048 points without any overlap. Any remaining data with less than 2048 points is discarded. For each type of samples, 80% are randomly selected for the training set and 20% for the test set. In the end, the training set comprises 9000 samples, and the test set comprises 2255 samples. Table 2 details the data distribution.

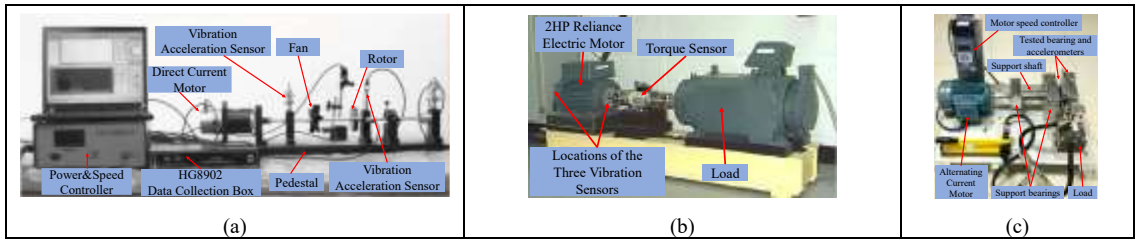

Fig.2. The experimental bench of each dataset: (a) PBR dataset, (b) CWRU dataset, (c) XJTU dataset.

Table 2. CWRU dataset

| Types | Marks | Diameter | Load(hp) | Label | Training | Test |
|---|---|---|---|---|---|---|
| Normal condition | NC | - | 0&1&2&3 | 0 | 662 | 166 |
| Inner raceway | F1 | 0.007 | 0&1&2&3 | 1 | 664 | 167 |
| Inner raceway | F2 | 0.014 | 0&1&2&3 | 2 | 665 | 167 |
| Inner raceway | F3 | 0.021 | 0&1&2&3 | 3 | 664 | 166 |
| Ball | F4 | 0.007 | 0&1&2&3 | 4 | 664 | 166 |
| Ball | F5 | 0.014 | 0&1&2&3 | 5 | 552 | 139 |
| Ball | F6 | 0.021 | 0&1&2&3 | 6 | 665 | 167 |
| Outer raceway | F7 | 0.007 | 0&1&2&3 | 7 | 1895 | 474 |
| Outer raceway | F8 | 0.014 | 0&1&2&3 | 8 | 664 | 166 |
| Outer raceway | F9 | 0.021 | 0&1&2&3 | 9 | 1905 | 477 |

Table 4. The parameters in TSViT

| Name | Description | Value |
|---|---|---|
| $L$ | the length of vibration signal | 2048 |
| $L_p$ | patch size<br>the width of the convolution kernel<br>the stride of the convolution | 32 |
| $m$ | the dimension of patch embedding<br>the number of output channels of the convolution | 192 |
| $h$ | the number of head in MSA | 8 |
| $d_{MLP}$ | the dimension of linear transformation in MLP | 768 |
| $B$ | the number of blocks in transformer encoder | 8 |
| $d_e$ | dropout probability in transformer encoder (MSA, MLP) | 0.1 |
| $d_p$ | dropout probability after position embedding | 0.1 |

Table 3. XJTU dataset

| Types | Marks | Label | Training | Test |
|---|---|---|---|---|
| Cage | F0 | 0 | 6822 | 1706 |
| Inner race | F1 | 1 | 19392 | 4848 |
| Outer race | F2 | 2 | 32486 | 8122 |
| Composite | F3 | 3 | 31948 | 7988 |

3) XJTU dataset. XJTU dataset is a dataset of rolling bearing faults and collected by the Institute of Design Science and Basic Component at Xi'an Jiaotong University [27]. It contains complete run-to-failure data of 15 rolling element bearings that were acquired by conducting many accelerated degradation experiments. Figure 2(c) presents the testbed for collecting XJTU dataset. The type of tested bearings is LDK UER204. Two accelerometers of type PCB 352C33 are mounted on the horizontal axis and vertical axis separately. The sampling frequency is set to 25.6 kHz. We selected the recorded data of Bearing 2_3, Bearing 3_1, Bearing 3_2, Bearing 3_4 for experiments. The selected data contains three single faults: cage, inner race, outer race, and one composite fault: inner race, ball, cage, outer race. It was processed with the same method as PBR and CWRU datasets. In the end, the training set comprises 90648 samples, and the test set comprises 22664 samples. Table 3 details the data distribution.

## 5 EXPERIMENTAL RESULTS AND EVALUATION

### 5.1 Results

Tables 4 lists the utilized parameters for TSViT model. The batch size during training was set to 32, 100 and 100 respectively based on the different sizes of PBR, CWRU and XJTU datasets. The learning rate was set to 0.0001. The trial was repeated 10 times under each condition to eliminate randomness.

Figures 3 to 8 depict the downward trend of the loss function and the upward trend of recognition accuracy in the PBR, CWRU, and XJTU datasets, respectively. Figures 3 (a) and (b) present the changes in losses during the first 35 epochs on the PBR training and test sets, respectively. The values of the loss function fluctuate due to random initialization, and the losses drop rapidly on both the training and test sets in the early stages of training. However, after 30 epochs, the boxes become flat, and the deviation range of the outliers gradually shrinks.

This indicates that the losses tend to stabilize. Both losses are close to 0 after 30 epochs. Figure 3 (c) presents the downward trends in the average loss throughout the 10 trials on both training and test sets. After the initial few epochs, the losses on both the training set and test set remain consistent. Both losses stabilize at 0 in the middle and late stages of training.

Figures 4 (a) and (b) present the changes of the two accuracies in the initial 35 epochs. The accuracies fluctuate and rise rapidly on both the training and test sets in the early stages

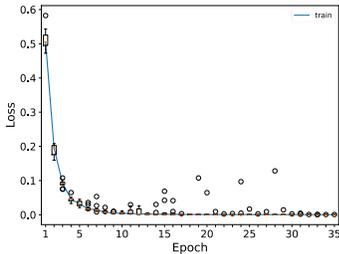
(a) downward trend on the training dataset

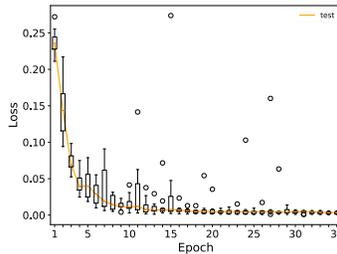
(b) downward trend on the test dataset

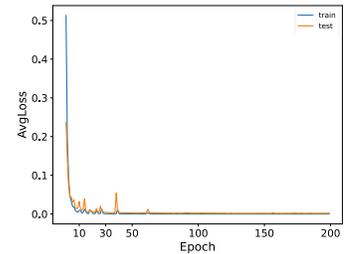
(c) downward trend in average loss over 10 trials

Fig. 3. The trend of loss changes on PBR dataset

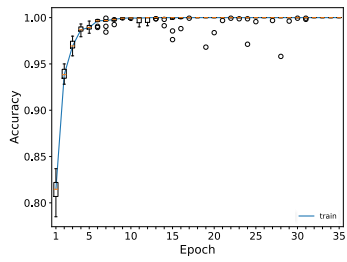
(a) upward trend on the training dataset

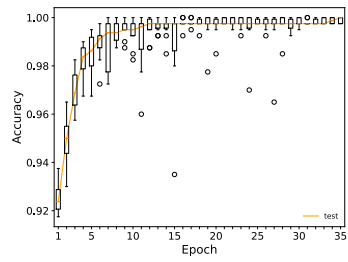
(b) upward trend on the test dataset

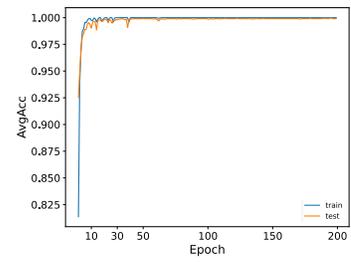
(c) upward trend in average accuracy over 10 trials

Fig. 4. The trend of accuracy changes on PBR dataset

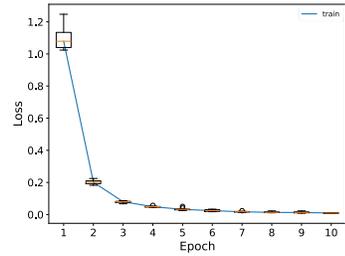
(a) downward trend on the training dataset

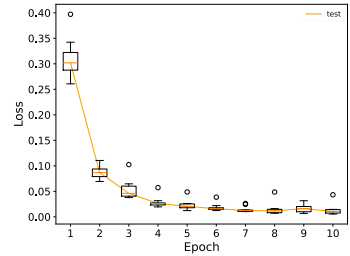
(b) downward trend on the test dataset

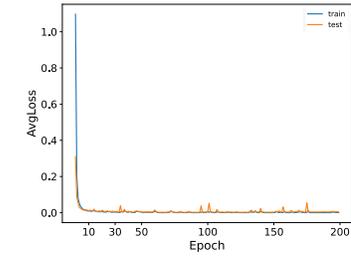
(c) downward trend in average loss over 10 trials

Fig. 5. The trend of loss changes on CWRU dataset

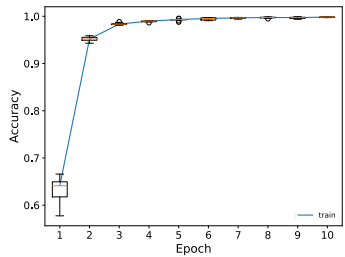
(a) upward trend on the training dataset

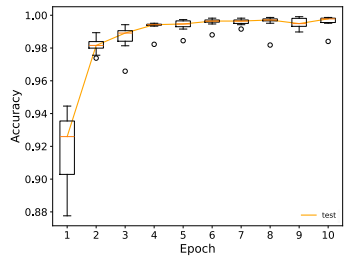
(b) upward trend on the test dataset

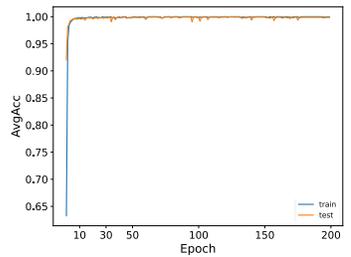
(c) upward trend in average accuracy over 10 trials

Fig. 6. The trend of accuracy changes on CWRU dataset

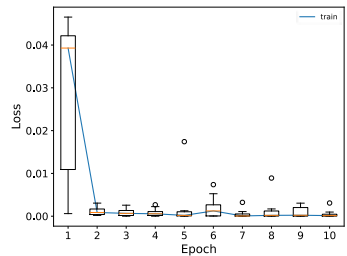
(a) downward trend on the training dataset

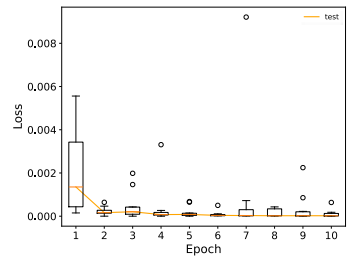
(b) downward trend on the test dataset

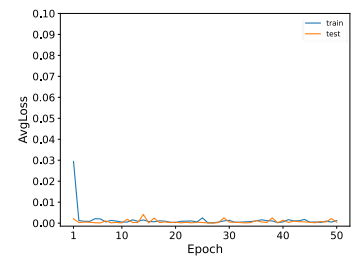
(c) downward trend in average loss over 10 trials

Fig. 7. The trend of loss changes on XJTU dataset

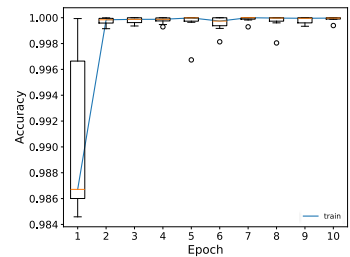
(a) upward trend on the training dataset

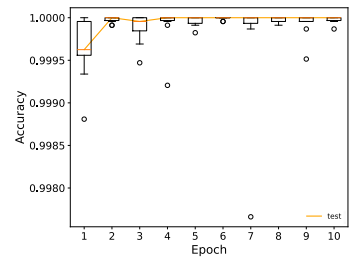
(b) upward trend on the test dataset

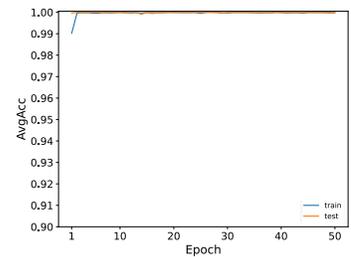
(c) upward trend in average accuracy over 10 trials

Fig. 8. The trend of accuracy changes on XJTU dataset

of training. However, after 30 epochs, the boxes become flat, the deviation range of the outliers gradually shrinks, indicating that both accuracies tend to stabilize. Both accuracies approach 100% after 30 epochs. Figure 4 (c) presents the upward trends in the average accuracy throughout the 10 trials on both the training and test sets. After the initial few epochs, the two accuracies remain consistent. Both accuracies stabilize at 100% in the middle and late stages of training, indicating that the model performs well and fits perfectly. The maximum accuracy of the optimal model (MaxAcc), the accuracy of the minimum optimal model (MinAcc), and the average accuracy of the optimal model (AvgAcc) are all 100% in 10 tests.

Table 5. Performance of TSViT

| Dataset | Accuracy | Precision | Recall | F1-score |
|---|---|---|---|---|
| PBR | 100% | 100% | 100% | 100% |
| CWRU | 99.99% | 99.99% | 99.99% | 99.99% |
| XJTU | 100% | 100% | 100% | 100% |

The results on CWRU dataset, as presented in Fig.5 and 6 are similar to that on PBR dataset. The two losses and accuracies gradually stabilize after only 10 epochs because CWRU dataset is much larger than PBR dataset. With 10 tests on the CWRU test set, the MaxAcc is 100%, the MinAcc is 99.96%, and the AvgAcc is 99.99%.

Figures 7 and 8 present the results on XJTU dataset. The two losses and accuracies gradually stabilize after only 1 epoch. With 10 tests on the XJTU test set, the MaxAcc, MinAcc, and AvgAcc are all 100%.

This study also evaluates the performance of TSViT with precision, recall, F1-score, which is presented in Table 5.

### 5.2 Noisy environment

In actual industrial scenarios, the collected vibration signals usually contain varying degrees of noise. Therefore, the anti-noise ability of TSViT is discussed in different noisy environments. Gaussian white noise with different signal-to-noise ratios (SNRs) is added to the original signal to simulate noise disturbance. The SNR is defined as Eq. (15).

$$SNR_{dB} = 10\log_{10}\left(\frac{P_{signal}}{P_{noise}}\right) \quad (15)$$

where $P_{signal}$ is the power of the signal and $P_{noise}$ is the power of the noise. If the SNR is 0 dB, the power of the noise is the same as the power of the original signal.

Noise signals with SNR ranging from 0dB to 10dB were added to the three datasets. The trial was repeated 10 times under each condition to eliminate randomness. Figure 9 presents the average accuracy with different SNRs on the three datasets throughout the 10 trails. Although the accuracy decreases as the noise signal increases, TSViT still performs well in noisy environments. The larger the dataset, the smaller the impact of the noise signal.

### 5.3 Hyperparameters

There are numerous structural hyperparameters utilized in TSViT. Different values for these hyperparameters may impact the performance of fault diagnosis. Therefore, it is essential to analyze them. We adjust the hyperparameters using the PBR,

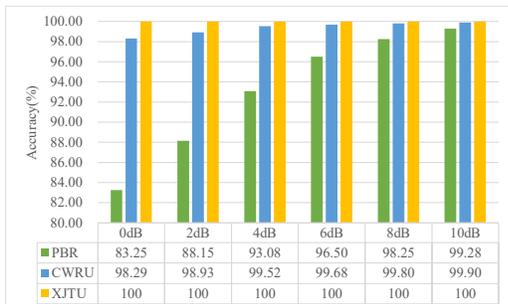

| | 0dB | 2dB | 4dB | 6dB | 8dB | 10dB |
|---|---|---|---|---|---|---|
| PBR | 83.25 | 88.15 | 93.08 | 96.50 | 98.25 | 99.28 |
| CWRU | 98.29 | 98.93 | 99.52 | 99.68 | 99.80 | 99.90 |
| XJTU | 100 | 100 | 100 | 100 | 100 | 100 |

Fig. 9. Accuracies on three noise-added test sets

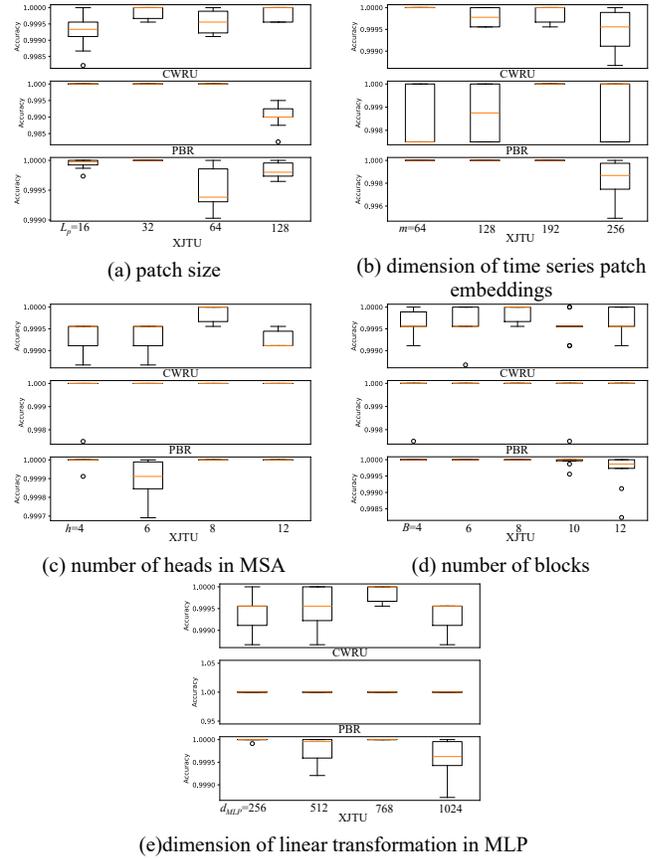

(a) patch size  
(b) dimension of time series patch embeddings  
(c) number of heads in MSA  
(d) number of blocks  
(e) dimension of linear transformation in MLP  

Fig.10. Comparison of the influence of different hyperparameters on accuracy based on three datasets

CWRU, and XJTU datasets. Table 6 presents the fault diagnosis results using TSViT model with various hyperparameter values under the same training condition. Figure 10 illustrates the results.

There are two crucial hyperparameters in the patch embedding phase: the size of a patch $L_p$ and the dimensions of the time series patch embedding $m$. The number of patches $n$ is inversely proportional to $L_p$, and this relationship influences the subsequent self-attention calculations. We assume that the length of signal samples $L$ must be evenly divisible by $L_p$ to ensure that a signal sample can be divided into integer patches. $L_p$ is varied between 16, 32, 64, and 128 to identify the optimal value for diagnostic testing. It can be seen from Table 6 that $L_p$ should not be excessively large. When $L_p$ is too large, the number of patches decreases, reducing the computational load of self-attention. However, this is not conducive to extracting features, and it also results in a decrease in the model's accuracy. It should not be excessively small either. If $L_p$ is too small, the overall model's computational load will increase exponentially, which is not favorable for model training and may lead to overfitting. Consequently, the accuracy of the model will also decrease. Figure 10(a) illustrates the results.

Table 6 presents that $m$ is related to the parameter quantity of the entire model and should not be excessively large. When $m$ is too large, there will be too many parameters, leading to overfitting. The average accuracy on the datasets is the highest when $m$ is set to 192 and $L_p$ is set to 32. Figure 10(b) illustrates

Table 6. Influence of the hyperparameters on the performance of TSViT

| | Hyperparameters | | | | | | | Flops | Params | Average accuracy | | |
|---|---|---|---|---|---|---|---|---|---|---|---|---|
| | $L_p$ | $h$ | $B$ | $d_{MLP}$ | $m$ | $pe$ | $dp$ | (M) | (M) | PBR | CWRU | XJTU |
| baseline | **32** | **8** | **8** | **768** | **192** | **Yes** | **Yes** | 309.88 | 2.39 | 100% | 99.99% | 100% |
| $L_p$ | 64 | | | | | | | 158.10 | 2.40 | 100% | 99.96% | 99.95% |
| | 128 | | | | | | | 82.21 | 2.42 | 99.03% | 99.98% | 99.98% |
| | 16 | | | | | | | 613.44 | 2.38 | 100% | 99.92% | 99.99% |
| $h$ | | 4 | | | | | | 309.88 | 2.39 | 99.98% | 99.93% | 100% |
| | | 6 | | | | | | 309.88 | 2.39 | 100% | 99.93% | 99.99% |
| | | 12 | | | | | | 309.88 | 2.39 | 100% | 99.93% | 100% |
| $B$ | | | 4 | | | | | 155.73 | 1.20 | 99.98% | 99.96% | 100% |
| | | | 6 | | | | | 232.81 | 1.79 | 100% | 99.96% | 100% |
| | | | 10 | | | | | 386.96 | 2.98 | 99.98% | 99.96% | 99.99% |
| | | | 12 | | | | | 464.03 | 3.57 | 100% | 99.96% | 99.97% |
| $d_{MLP}$ | | | | 256 | | | | 105.41 | 0.81 | 100% | 99.94% | 100% |
| | | | | 512 | | | | 207.65 | 1.60 | 100% | 99.96% | 99.98% |
| | | | | 1024 | | | | 412.12 | 3.18 | 100% | 99.93% | 99.95% |
| $m$ | | | | | 64 | | | 103.29 | 0.80 | 99.85% | 100% | 100% |
| | | | | | 128 | | | 206.59 | 1.59 | 99.88% | 99.98% | 100% |
| | | | | | 256 | | | 413.18 | 3.18 | 99.90% | 99.94% | 99.84% |
| $pe$&$dp$ | | | | | | Yes | No | 309.88 | 2.39 | 99.80% | 99.96% | 100% |
| | | | | | | No | Yes | 309.88 | 2.39 | 99.98% | 99.97% | 99.97% |

Note: $pe$ is the abbreviation of position embedding and $dp$ represents the dropout after $pe$.

the results.

The model's performance is slightly affected by the position embedding. Experimental results indicate a 0.02% decrease in average accuracy on the three datasets. Additionally, the dropout following the position embedding also plays a role in influencing performance. Without dropout, the model tends to overfit, resulting in a decrease in average accuracy on the three datasets.

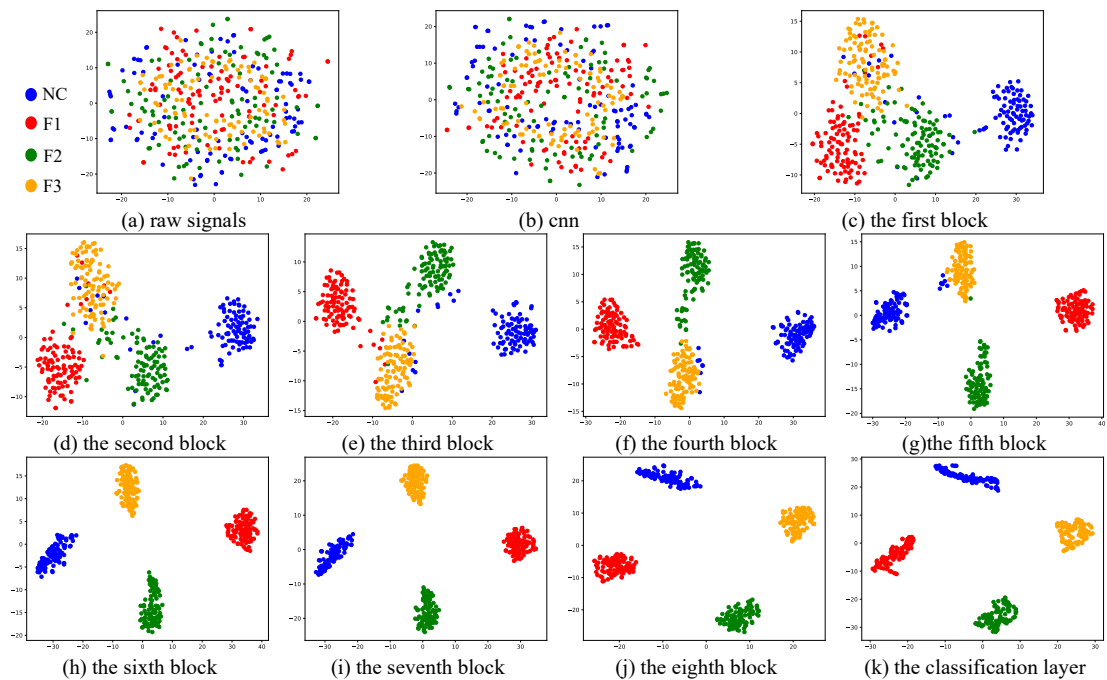

Fig.11. Feature visualiztion in different TSViT layers through t-SNE on the PBR test set

(a) raw signals
(b) cnn
(c) the first block
(d) the second block
(e) the third block
(f) the fourth block
(g) the fifth block
(h) the sixth block
(i) the seventh block
(j) the eighth block
(k) the classification layer

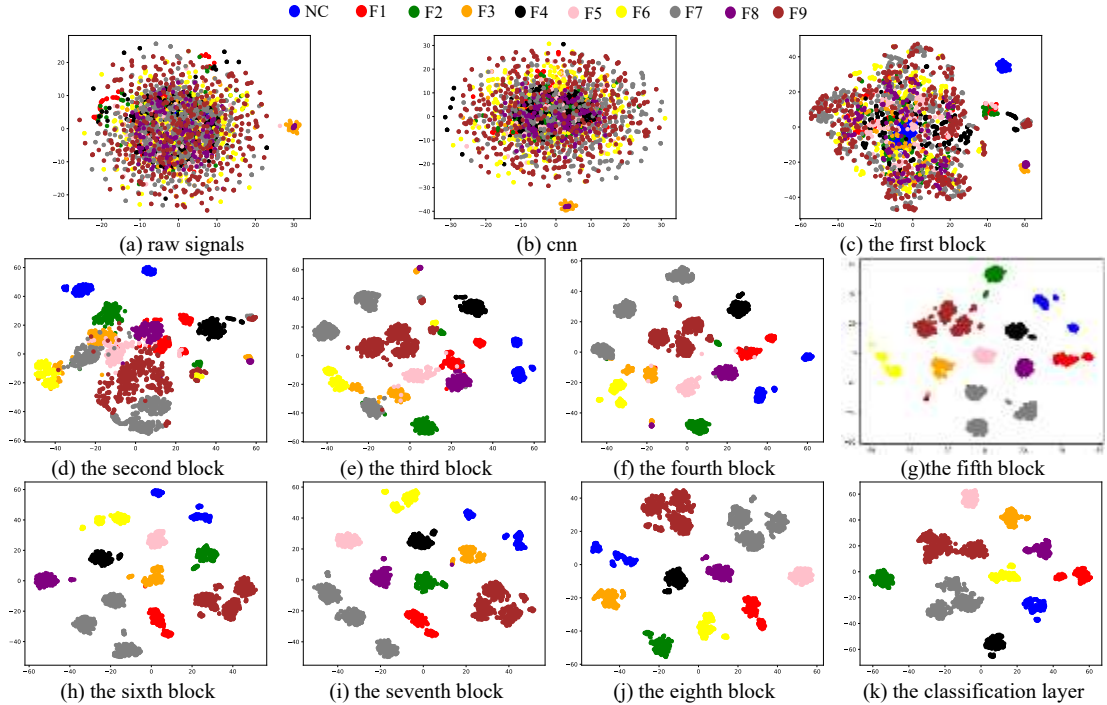
Fig.12. Feature visualiztion in different TSViT layers through t-SNE on the CWRU test set

Within the transformer encoder layer, three crucial hyperparameters are the number of heads in MSA $h$, the number of blocks $B$, and the dimension of linear transformation in MLP $d_{MLP}$. As indicated by Table 6, the computation load and parameter quantity remain constant, and the accuracy rises with an increase in $h$. However, when $h$ becomes excessively large, the feature subspace shrinks, hindering feature extraction. Figure 10(c) illustrates the results.

Table 6 further indicates that the average accuracy increases with the enlargement of $B$ or $d_{MLP}$. However, this increase in $B$ or $d_{MLP}$ also results in a significant rise in the computational load and parameter quantity of the entire model. When these values become excessively large, overfitting may occur, leading to a decline in accuracy. Figure 10(d) and (e) illustrates the results.

5.4 Visualization of feature vectors

The feature vector distribution within the embedding space

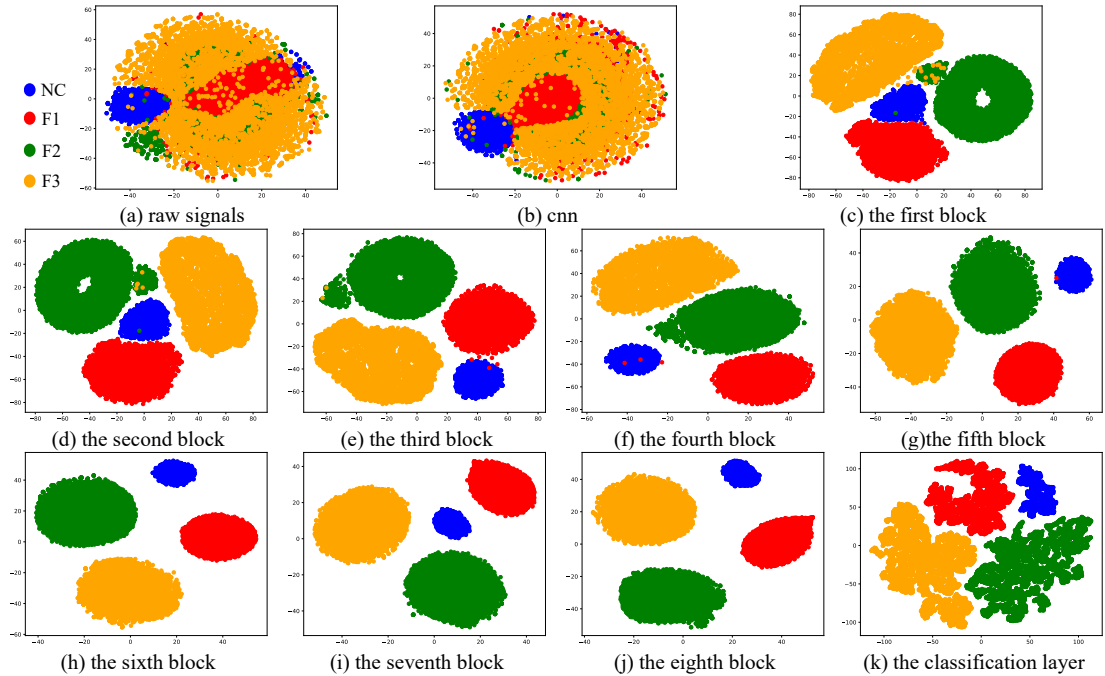
Fig.13. Feature visualiztion in different TSViT layers through t-SNE on the XJTU test set

serves as a meaningful indicator to assess a model's generalization capability [28]. Here, we employed t-distributed Stochastic Neighbor Embedding (t-SNE) [29] to visualize the feature vectors extracted by different layers. This visualization reflects the feature learning and classification processes of TSViT.

We employed t-SNE to visualize two complete test sets, aiming to validate TSViT's effectiveness. The feature vectors include patch embeddings and class tokens. Figures 11, 12, and 13 depict the visual results of different layers of TSViT through t-SNE on PBR, CWRU, and XJTU test sets respectively. As observed in Fig.11 (a), Fig.12(a) and Fig.13 (a), the feature vectors of the raw vibration signals across different health states show substantial mixing and overlap, preventing clear differentiation between states. This phenomenon persists even within the CNN layer. However, surprisingly, there is a notable improvement observed on PBR and XJTU test sets in the first block. It shows that the feature vectors in this block are distinct and easily distinguishable already. A similar situation is observed in the second block for the CWRU test set. Failure categories are entirely distinct in the 6th block for PBR and XJTU test sets. In the 7th block, only a few F9 samples fall within the decision boundary of F3. Both F3 (inner raceway) and F9 (outer raceway) have fault diameters of 0.021 inches. Similarly, in the 8th block of the CWRU test set, failure categories are completely separated as well. These visual results illustrate the robustness of TSViT in features extraction.

5.5 Comparative analysis

We selected three deep learning models for comparing and validating the TSViT model for fault diagnosis. These three models include Deep Convolutional Neural Networks with wide first-layer kernels (WDCNN) [30], Long Short-Term Memory (LSTM) [31], and CNN-LSTM [32]. These three models are widely applied in fault diagnosis. The details are presented in Table 7.

WDCNN consists of 5 convolution and pooling layers and 2 linear layers. The convolution kernel in its first layer has a wide width to extract features and suppress high-frequency noises. WDCNN can achieve fault diagnosis based on raw vibration signals. LSTM is a specialized type of RNN designed to address the challenges of gradient vanishing and explosion encountered during training with long sequences [31]. The length of each signal sample is 2048, which is considered too long for LSTM. This length can pose challenges as subsequent LSTM units may find it difficult to capture information from previous units. To address this issue, we reshaped the signal samples into 64x32 matrices. CNN-LSTM utilizes CNN to extract local features from vibration signals, and LSTM to learn the temporal dependencies among these features. The above three models are widely applied in fault diagnosis based on vibration signals.

In our comparative analysis, we also evaluate the performance of our approach alongside the achievements of other works which have applied transformers to fault diagnosis. This includes Time Series Transformer (TST) [28], Efficient Convolutional Transformer (ECTN) [33], Integrated ViT [20], and MCF-1DViT [19]. ECTN and Integrated ViT transform vibration signals into time-frequency representation maps, while MCF-1DViT processes signal samples with a length of 1024. These methods employ different mechanisms compared to TSViT model. Therefore, we directly use their experimental results from the literature for comparison.

Table 8 presents the comparison of results with other methods, highlighting the superiority of TSViT in fault diagnosis. TSViT achieves the highest accuracy among these methods without any preprocessing for vibration signals.

Table 7. The detailed structure of the comparison models

| Stage | WDCNN | LSTM | | CNN-LSTM |
|---|---|---|---|---|
| 1 | Convolution (channels=16, kernel size=64, stride=16), Batchnorm, Relu, Maxpooling(2) | Reshape(64×32), Linear(32,192) | | Convolution (channels=32, kernel size=64, stride=32), Elu, Maxpool(4,2) |
| 2 | Convolution (channels=32, kernel size=3, stride=1), Batchnorm, Relu, Maxpool(2) | LSTM(192,192) Dropout(0.1) | ×8 | Convolution (channels=32, kernel size=5, stride=1), Elu, Maxpool(4,2) |
| 3 | Convolution (channels=64, kernel size=3, stride=1), Batchnorm, Relu, Maxpool(2) | ×3 | Linear(128, num_class) | Convolution (channels=64, kernel size=3, stride=1), Elu, Avgpool() |
| 4 | Linear(192,100) Batchnorm Linear(100,num_class) | | | Linear(64,64), Elu LSTM(64,32), Linear(32,num_class) |
| | FLOPs=1.76M Params=0.06M | FLOPs=306.71M Params=2.39M | | FLOPs=2.25M Params=0.06M |

Table 8. Comparison of results with other methods

| CWRU dataset | | |
| --- | --- | --- |
| Methods | Preprocessing | Accuracy |
| WDCNN | No | 90.11% |
| LSTM | No | 99.43% |
| CNN-LSTM | No | 98.58% |
| TST | No | 99.91% |
| MCF-1DViT | No | 99.83% |
| Integrated ViT | DWT+CWT | 99.87% |
| ECTN | STFT | 99.62% |
| TSViT | No | 99.99% |

| PBR dataset | | |
| --- | --- | --- |
| Methods | Preprocessing | Accuracy |
| WDCNN | No | 73.50% |
| LSTM | No | 98.30% |
| CNN-LSTM | No | 93.68% |
| TST | No | 99.98% |
| TSViT | No | 100% |

| XJTU dataset | | |
| --- | --- | --- |
| Methods | Preprocessing | Accuracy |
| WDCNN | No | 99.68% |
| LSTM | No | 100% |
| CNN-LSTM | No | 99.98% |
| TST | No | 100% |
| TSViT | No | 100% |

Notably, transformer-based models outperform other approaches overall. Since XJTU dataset is much larger than the other two datasets, all the models perform best on it. WDCNN is overfitting on PBR and CWRU datasets. LSTM and TST models also perform well. However, the efficiency of LSTM is low due to the poor parallelism. Since the sequence length in the TST model is too long, the amount of calculation is too large, and the efficiency is low.

6 CONCLUSIONS AND FUTURE WORK

Vision transformer tends to be highly promising for fault diagnosis. This study proposed TSViT model for fault diagnosis of rotating machinery, which can process raw vibration signals without any preprocessing. It seamlessly integrates transformer and CNN, addressing the limited receptive field issue of convolution kernels while effectively capturing long-term dependencies in vibration signals. The multi-head self-attention mechanism enables the model to capture pertinent information across various representation subspaces, enhancing the interpretability of the diagnostic model. The experimental results on PBR, CWRU and XJTU datasets validate the effectiveness of TSViT under various working conditions, including different loads and speeds. This paper also analyzes the influence of its hyperparameters on model performance, computational complexity, and overall parameter quantity through experiments. This analysis provides valuable insights for researchers, facilitating the adoption of vision transformer in their work. The comparative experiments with other methods on the three distinct datasets demonstrate the superiority of TSViT model. The findings from TSViT illustrate the effective application of vision transformers in analyzing time series vibration signals for industrial fault diagnosis. This suggests that deep learning-based fault diagnosis in industry is also expected to be unified under the transformer structure.

The advent of Industry 4.0 has led to the generation of massive datasets with the help of IIoT. The transformer model has demonstrated remarkable scalability to handle large models and big data. However, challenges arise due to the scarcity of fault samples in real industrial settings, it is essential to investigate effective applications of transformers for small-sample datasets.

**Acknowledgments:** We would like to express our sincere gratitude to Professor Weishan Zhang, Professor Chenglin Wen and Professor Chen Yu for their valuable guidance and support throughout this research.